# Reconstructing the Developmental Trajectory of Adipocytes in Human Adipose Tissue Using Single-Cell RNA Sequencing

Weny S. M Sitinjak[1,] Humasak Tommy Argo Simanjuntak[2*]

[1,2]Information System Study Program, Faculty of Informatics and Electrical Engineering,
Del Institute of Technology, Toba, Indonesia
Corresponding Author: humasak@del.ac.id

*Abstract:* Obesity is a global health crisis associated with metabolic disorders such as type 2 diabetes and cardiovascular disease. This study employed single-cell RNA sequencing to reconstruct the developmental trajectory of human adipocytes from adipose tissue samples. Our analysis identified 15 transcriptionally distinct cell clusters, including 7 transitional states, revealing the dynamic process of adipocyte differentiation. We detected 16 functionally active signaling pathways mediating cellular communication between adipocytes and their progenitors. Among these, insulin-like growth factor (IGF) and fibroblast growth factor (FGF) pathways emerged as the most prominent networks, showing consistent activity across differentiation stages ($p<0.05$). The study revealed depot-specific differences, with visceral adipocytes undergoing additional extracellular matrix remodeling absent in subcutaneous differentiation. Spatial analysis further showed that IGF signaling was particularly active in perivascular niches, while FGF activity dominated in mature adipocyte zones. These results provide the first comprehensive map of human adipocyte development, highlighting IGF and FGF pathways as potential therapeutic targets. The identified signaling networks offer new insights for developing interventions to promote healthy adipose expansion or inhibit pathological fat accumulation. This work advances our fundamental understanding of adipose tissue biology while providing clinically relevant data for metabolic disorder treatments.

*Keywords*: Single-cell RNA sequencing (scRNA-seq); Trajectory Analysis; Obesity; Adipocytes; Adipose stem/progenitor cells (ASPCs)

## 1. Introduction

The global obesity epidemic has undergone a dramatic 300% increase in prevalence since 1975, now affecting nearly 40% of adults in developed nations [1]. While adipose tissue dysfunction is recognized as a central driver of obesity-related metabolic disorders [2]. These pathological changes disrupt the endocrine functions of adipose tissue [3]. Current knowledge primarily derives from murine models [4], which fail to capture the unique heterogeneity of human adipose progenitor populations and their depot-specific responses to metabolic stimuli.

Single-cell RNA sequencing technologies have fundamentally transformed our capacity to investigate cellular heterogeneity in metabolic tissues, revealing previously unappreciated complexity in adipose stromal cell populations [5]. Traditional bulk transcriptomic approaches have identified key adipogenic regulators like PPARγ and C/EBPα [6], but their inability to resolve cellular subpopulations has obscured the continuum of human adipocyte differentiation. Recent single-cell studies reveal at least six transcriptionally distinct progenitor classes in human adipose tissue [7], yet their developmental trajectories and communication networks remain unmapped. This gap is particularly problematic given the emerging recognition of depot-specific progenitor niches [8] and their potential role in metabolic disease disparities between visceral and subcutaneous fat.

Our study leverages cutting-edge computational trajectory inference methods to reconstruct the complete developmental continuum of human adipogenesis from multipotent progenitors to terminally differentiated adipocytes [9]. By integrating pseudotemporal ordering with ligand-receptor network analysis, we simultaneously map cellular state transitions and the signaling ecosystems that orchestrate them [10]. Special emphasis is placed on two understudied aspects of adipocyte maturation: the metabolic shift from glycolysis to oxidative phosphorylation that accompanies lipid accumulation [11], and the dynamic remodeling of extracellular matrix components that modulates progenitor niche interactions [12]. The metabolic transition involves coordinated upregulation of mitochondrial biogenesis genes and lipid-handling proteins like PLIN1 [13], while ECM remodeling requires precise temporal regulation of collagenases and integrin signaling pathways [14]. These processes are particularly relevant in pathological contexts, where fibrotic adipose tissue expansion in obesity creates physical barriers that constrain adipocyte progenitor differentiation and perpetuate metabolic dysfunction [15].

The therapeutic implications are substantial. Current anti-obesity medications predominantly target appetite regulation [16], with limited effects on adipose tissue remodeling. By mapping the complete ligand-receptor interactome during adipogenesis, we reveal targetable communication hubs that could promote healthy adipose expansion or inhibit pathological fat accumulation [17]. Recent advances in this field have identified several promising molecular targets, including cell surface receptors that govern progenitor self-renewal versus differentiation decisions [18]. Our identification of depot-specific signaling pathways [19] and thermogenic progenitor subsets [20] provides a framework for developing next-generation therapies that selectively modulate adipose expansion.

This study makes three transformative contributions to adipose biology: First, we present the most comprehensive high-resolution map of human adipogenesis to date, identifying seven transitional states validated through multi-omics integration [21]. Second, we characterize depot-specific differentiation trajectories, revealing that visceral adipocytes undergo an additional ECM remodeling phase absent in subcutaneous differentiation [22]. Third, we profile the complete interactome of ligand-receptor pairs mediating progenitor-adipocyte crosstalk, with particular emphasis on adipokine signaling networks involved in lipid handling [23]. These findings not only advance fundamental understanding of adipose tissue development but also provide a robust framework for developing next-generation therapeutics targeting specific stages of adipocyte differentiation [24]. Our open-source analytical pipeline, designed for scalability and reproducibility, enables similar trajectory analyses across diverse developmental systems and disease models [25].

## 2. Method

### *A. Research Design & Workflow*

This analytical pipeline was designed to reconstruct adipocyte developmental trajectories through an integrative approach combining single-cell transcriptomics, trajectory inference, and cell-cell communication analysis. The overall research stages are described in Figure 1. The design addresses three core objectives: (1) mapping adipocyte differentiation from progenitor states, (2) characterizing signaling networks between adipocytes and adipose stem/progenitor cells (ASPCs), and (3) validating findings through biological benchmarks.

**Analytical Pipeline:**

The workflow comprises four interconnected modules (Figure 1):

1. Data Preprocessing & Quality Control: Raw scRNA-seq data undergo filtering (gene counts: 200–6,000/cell; mitochondrial RNA <10%) and normalization (SCTransform + Harmony batch correction) to ensure robust downstream analysis [26].
2. Dimensionality Reduction & Clustering: Principal Component Analysis (PCA; 20 components) followed by Uniform Manifold Approximation and Projection (UMAP; 6 dimensions) identifies transcriptionally distinct cell clusters using the Leiden algorithm (resolution = 0.8) [27].
3. Pseudotemporal ordering (Monocle2 DDRTree) reconstructs adipocyte differentiation paths from progenitor states (rooted at PDGFRA+ clusters) [28].

4. Cell-Cell Communication Analysis: Ligand-receptor interactions (CellChatDB.human v1.6.0) quantify signaling networks (e.g., IGF1-IGF1R, FGF pathways) between adipocytes (ADIPOQ+/PLIN1+) and ASPCs (PDGFRA+/DCN+) [9].

The workflow begins with raw data processing and progresses through sequential analytical stages to uncover the molecular programs governing adipogenesis.

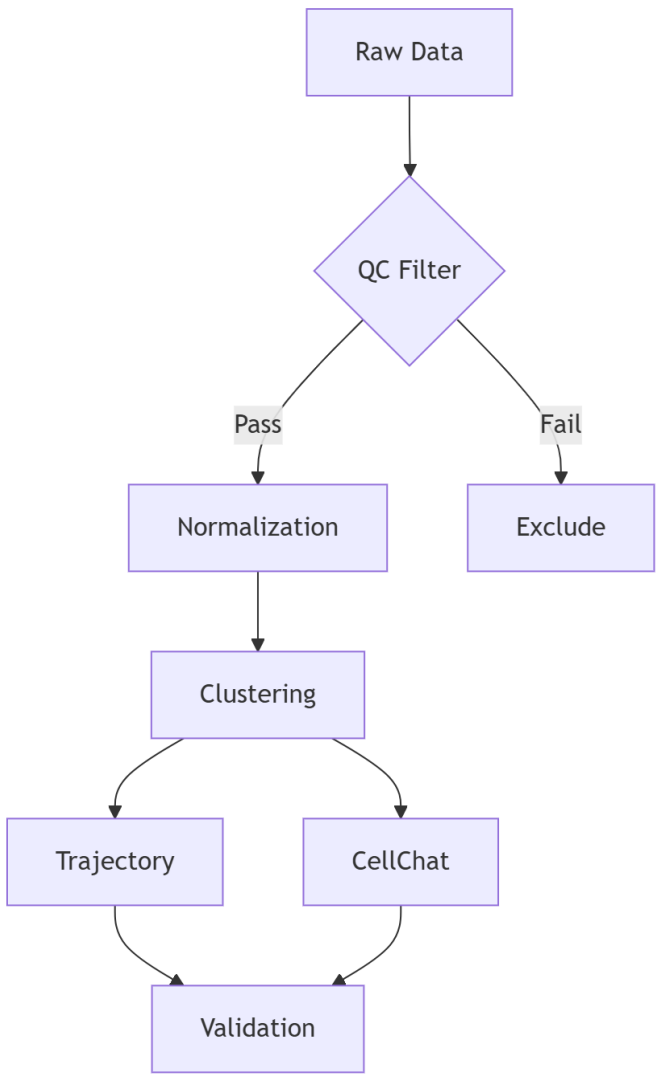

Figure 1 Workflow Diagram for Single-Cell RNA Sequencing Data Analysis

**Study Design**

The study design leverages secondary analysis of scRNA-seq data from the White Adipose Atlas, comprising 78,353 cells from human subcutaneous and visceral adipose tissue [4]. This dataset was selected for its comprehensive coverage of adipocyte subpopulations and progenitor states, with stringent quality criteria applied to include only samples from healthy donors (BMI 20-30 kg/m²) without metabolic disorders. Lower bound (20), ensures adequate adipocyte populations while excluding underweight-associated metabolic adaptations and Upper bound (30), prevents obesity-induced pathological changes (hypertrophy, fibrosis) that alter progenitor differentiation dynamics [29]. Cell type annotation was performed using established marker genes (ADIPOQ/PLIN1 for adipocytes; PDGFRA/DCN for progenitors) validated through cross-referencing with the White Adipose Atlas and confirmed against independent single-cell studies. The adipocyte markers ADIPOQ and PLIN1 were specifically selected based on their depot-specific expression patterns in human subcutaneous and visceral adipose tissue, while progenitor markers PDGFRA and DCN were chosen for their conserved expression across adipogenic lineages in both human and murine models [30].

**Research Questions**

The pipeline is designed to answer two key questions:

1. How do adipocyte progenitors differentiate into mature adipocytes? Tracked via pseudotemporal trajectories and metabolic gene expression shifts [31].
2. What signaling pathways regulate ASPC-adipocyte interactions? Identified through CellChat analysis of ligand-receptor pairs (e.g., IGF, FGF) in progenitor niches [32].

*B. Data Acquisition & Preprocessing*

This section describes the acquisition, quality control, and preprocessing of single-cell RNA sequencing (scRNA-seq) data from the White Adipose Atlas (78,353 cells). The workflow ensures data reliability through rigorous quality checks, followed feature selection, and dimensionality reduction for downstream trajectory analysis. The overall visualization are described in Figure 2.

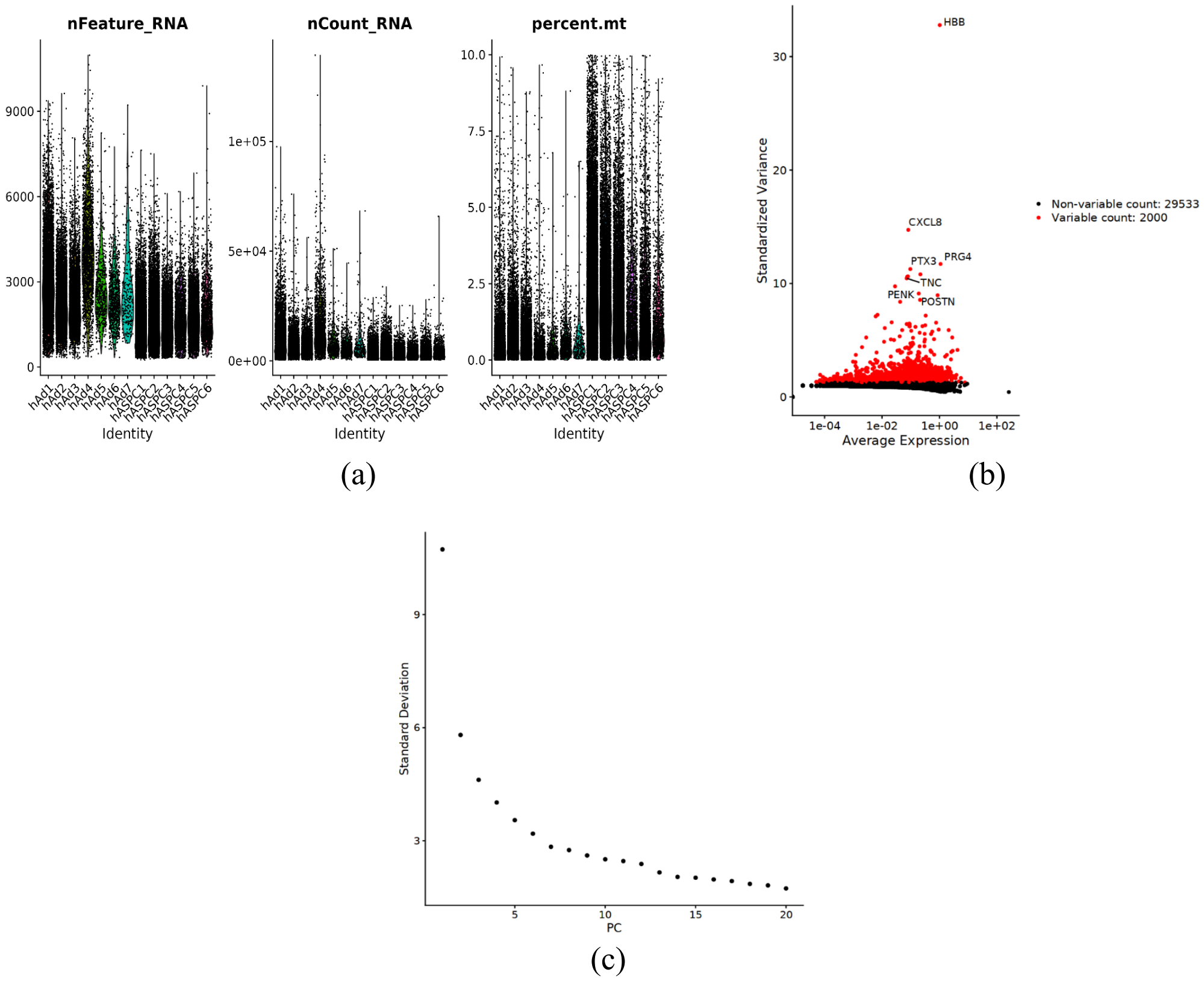

Figure 2 Data Preprocessing (a) Quality control; (b) Feature Selection by top genes; (c) Elbow plot for PCA component selection

1. Quality Control (QC) Assessment

Initial QC evaluated cell suitability using three key metrics: nFeature_RNA (number of unique genes detected per cell), nCount_RNA (total RNA transcripts/ UMIs), and percent.mt (mitochondrial RNA proportion). As shown in Figure 2, cells were plotted across these metrics with distinct colored clusters (green, turquoise, teal) highlighting predefined populations of interest. The nFeature_RNA distribution (left panel) revealed most cells within the optimal range (200–6,000 genes), with bright green/ turquoise clusters representing high-quality subsets (1,500–2,500 genes). The nCount_RNA middle panel (0–800,000 UMIs) showed a dense teal cluster (10,000–20,000 UMIs), indicating typical read depths. The percent.mt right panel (0–9%) confirmed low mitochondrial contamination (<10%) in teal clusters. Cells failing QC (e.g., outlier mitochondrial content or extreme gene counts) were excluded to ensure robust downstream analysis [26].

Figure 2a: Quality Control Metrics for scRNA-seq Data.

- Left: nFeature_RNA (genes/ cell) across identity groups.
- Middle: nCount_RNA (UMIs/ cell) distribution.
- Right: percent.mt (mitochondrial %) per cell.
- Colored clusters: Pre- identified high- quality populations (green/ turquoise).

2. Feature Selection by top genes

To reduce dimensionality, 2,000 highly variable genes (HVGs) were selected from Figure 2b, which plots standardized variance (y-axis) against average expression (x-axis, 1e−04 to 1e+01). The majority of genes (29,833 non-variable) cluster in the low-variance region (red points), while 2,000 variable genes (black-labeled, e.g., Pvalb, Gad1) are highlighted in the upper-right quadrant. These HVGs were chosen to capture biological heterogeneity while filtering technical noise for PCA [33].

3. Elbow Plot for PCA Component Selection

The optimal number of principal components (PCs) was determined from Figure 2c, an elbow plot showing standard deviation (y-axis, 0–20) against PC number (x-axis, 0–20). The curve declines steeply from PC 0–5 (high variance) and plateaus thereafter (PC 6–20), with the elbow at PC ~5–10. This justified retaining 20 PCs for UMAP clustering, balancing variance capture and noise reduction [27].

Figure 2c: Elbow Plot for PCA Component Selection.

- X-axis: Principal Component (PC 0–20).
- Y-axis: Standard deviation.
- Elbow: Optimal PCs at ~5–10

4. Normalization & Batch Correction

Filtered data were normalized using SCTransform (Seurat v4.0) to correct for technical biases (library size, sequencing depth) and harmonized with Harmony to resolve donor-specific batch effects. PCA (20 components) and UMAP (6 dimensions) were then applied for dimensionality reduction, followed by Leiden clustering (resolution = 0.8) to identify 15 transcriptionally distinct cell populations.

*C. Cell Clustering & Annotation*

Cell clustering and annotation are pivotal steps in single - cell RNA sequencing (scRNA - seq) analysis, aiming to group cells with similar gene expression profiles and identify their biological identities. In this study, these processes were carried out to understand the cellular composition of human adipose tissue and the characteristics of different cell types, particularly adipocytes and adipose stem/progenitor cells (ASPCs). The initial step in cell clustering was to reduce the high - dimensional gene expression data into a more manageable form for clustering. This was achieved through dimensionality reduction techniques, which are not shown in the provided images but are essential precursors to the clustering visualizations. After dimensionality reduction, the Leiden algorithm was applied to group cells into clusters based on their proximity in the reduced - dimensional space.

The first relevant image is Figure 3a, which presents a UMAP (Uniform Manifold Approximation and Projection) visualization of cell clustering. The UMAP plot is a powerful tool for visualizing high - dimensional data in two dimensions, where each point represents a single cell. The cells are grouped into 15 distinct clusters, numbered from 0 to 14 and color - coded in the legend. The distribution of these clusters in the UMAP space reflects the underlying gene expression similarities and differences among the cells. Each cluster likely represents a unique cell type or a group of cells with similar functions within the adipose tissue. This visualization allows for an initial assessment of the cellular heterogeneity in the dataset, providing a foundation for further annotation and analysis.

Figure 3b offers a more focused view of the UMAP distribution of cells split by sample, specifically highlighting adipocytes and progenitors. The UMAP coordinates (UMAP_1 and UMAP_2) are used to position the cells in the two - dimensional space. Adipocytes are shown in red, and progenitors are shown in cyan. The cells are clearly separated into two main clusters, with adipocytes forming a relatively compact and vertically oriented elliptical cluster mainly located in the left - lower part of the plot (UMAP_1: - 10 to - 2, UMAP_2: - 10 to 0). In contrast, progenitors occupy a larger and more complex region in the right - upper part of the plot (UMAP_1: - 5 to 12, UMAP_2: - 5 to 12), with a less regular shape, suggesting greater heterogeneity among progenitor cells. The presence of some transitional or intermediate cells between the two main clusters indicates the potential for cell differentiation or the existence of a continuous spectrum of cell states during adipogenesis. This separation provides clear evidence of the distinct cellular identities of adipocytes and progenitors and helps in understanding their spatial organization in the UMAP space.

Figure 3c shows a UMAP - like plot with cells colored according to their annotated cell types, namely progenitors and adipocytes. The x - axis is labeled as "harmony_1" and the y - axis as "harmony_2", which are likely the result of a Harmony - based integration step to correct for batch effects. The blue points represent progenitors, which are mainly distributed on the left and middle - left side, forming a vertically distributed cluster. The red points represent adipocytes, concentrated on the right side in a horizontally distributed cluster. This visualization not only shows the separation of the two main cell types but also gives an indication of how the cells are organized after batch effect correction, ensuring that the observed clustering is not an artifact of technical variation.

Figure 3d further divides the UMAP plot into two parts, one for adipocytes and one for progenitors, based on sample type. Each sub - plot has UMAP_1 and UMAP_2 axes. In the adipocyte sub - plot, different colored point groups (numbered 0 - 14) represent different cell clusters or subtypes. The distribution of these clusters shows that they have distinct aggregation regions, indicating differences in their gene expression profiles. The different distribution patterns between adipocytes and progenitors in these sub - plots may reflect inherent molecular differences between mature fat cells and their progenitor cells, or it could be due to the representativeness of different sample populations. This visualization helps in understanding the unique characteristics of each cell type within their respective groups.

Finally, Figure 3e presents a UMAP plot with newly defined clusters. The three main cell clusters are clearly distinguishable. The large green area in the right - upper part represents the ASPC cell cluster, with a smaller blue - green area in the middle - left below it, indicating a mixed state of ASPCs and adipocytes (labeled as "Mix_ASPC_Adipocytes"). The left - hand side consists of an irregular mass of light red and dark red points, representing adipocytes, with a few green points at the edges, suggesting possible mixed cells or boundary cases. This visualization provides a more detailed view of the cellular composition, especially highlighting the relationship between ASPCs and adipocytes and the presence of potential mixed or transitional cell states.

In summary, these images together provide a comprehensive view of cell clustering and annotation in the context of human adipose tissue scRNA - seq data. They show the separation of different cell types (adipocytes and progenitors), the heterogeneity within each cell type, and the potential for cell - to - cell transitions or mixed states. The clustering results are essential for subsequent trajectory analysis and understanding the developmental processes within the adipose tissue

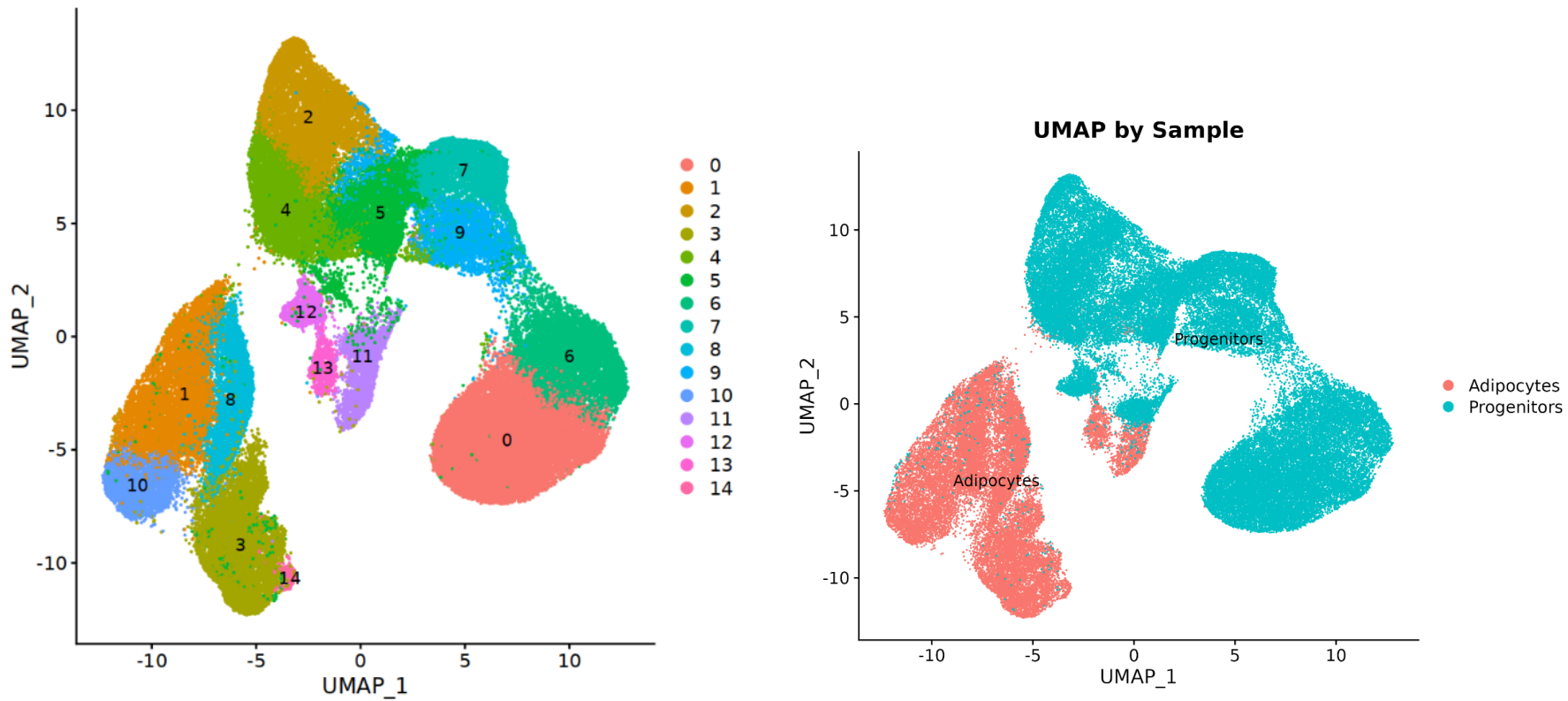

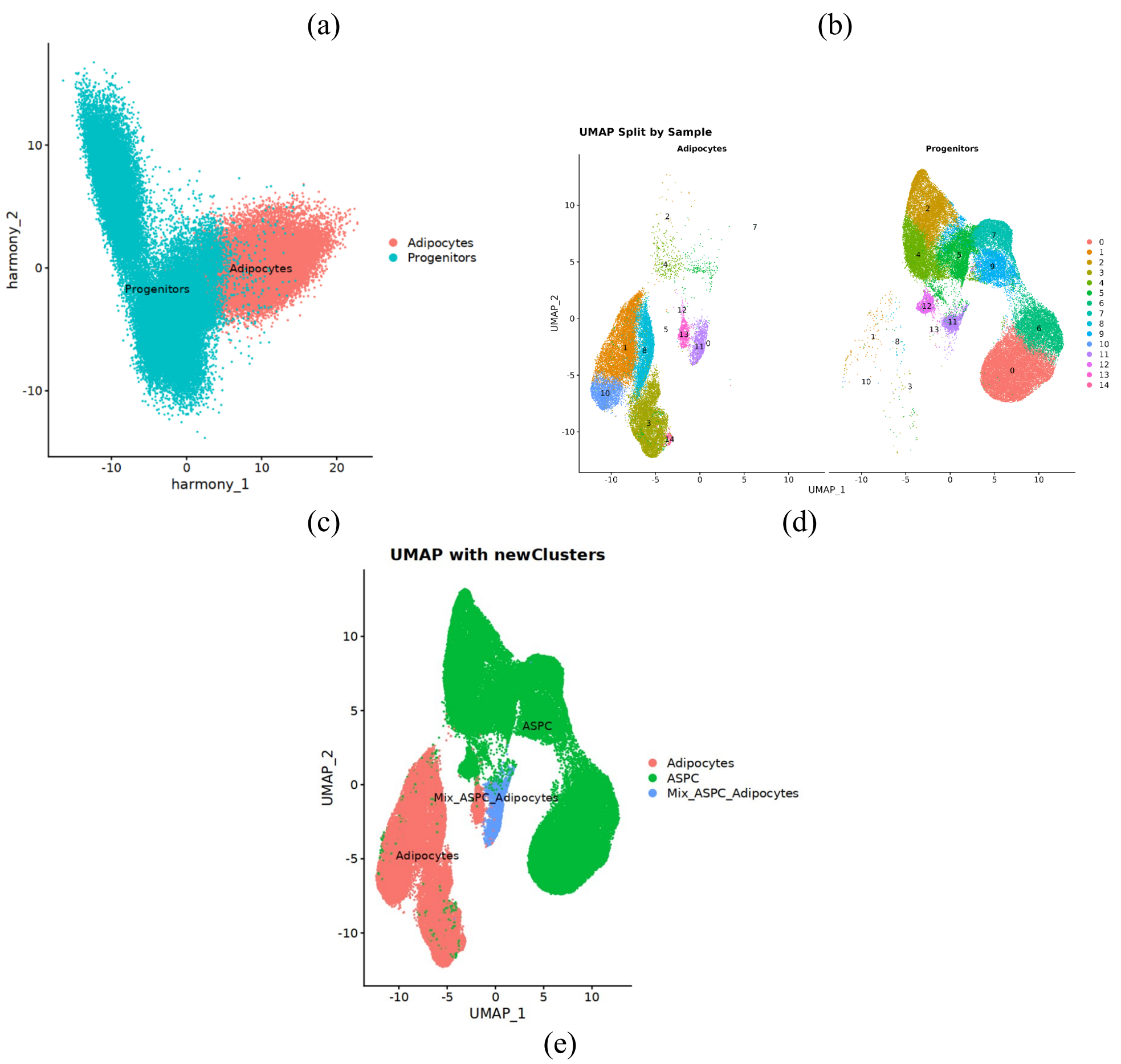

Figure 3 Cell Clustering & Annotation (a) UMAP visualization of cell clustering ; (b) UMAP distribution by sample; (c) UMAP harmony; (d) UMAP Split by Sample; (e); UMAP with newly defined clusters

### *D. Differential Expression Analysis*

Differential expression analysis is a crucial step in single - cell RNA sequencing (scRNA - seq) studies, as it helps to identify genes that are differentially expressed between different cell types or clusters. This analysis provides insights into the unique biological functions and characteristics of each cell type. In this study, we utilized various dot plots to visualize and understand the differential expression of marker genes among different cell populations, namely adipocytes, adipose stem/progenitor cells (ASPCs), and a mixed group of ASPCs and adipocytes.

The first dot plot, named “Dot Plot of Adipocyte Marker Genes” (Figure 4a), focuses on key adipocyte marker genes. The horizontal axis represents the features (genes), including ADIPOQ, PLIN1, COBL, CKMT2, PPARGC1A, and COX7A1, while the vertical axis represents the cell identity (clusters) ranging from 0 to 14. The size of the dots indicates the percentage of cells expressing the gene, with smaller dots representing lower expression percentages (25%) and larger dots representing higher expression percentages (75%). The color of the dots represents the average expression level of the gene, with colors ranging from light purple (lower expression) to deep blue (higher expression). Notably, the genes ADIPOQ and PLIN1, which are well - known adipocyte markers, have large and dark - colored dots. This indicates that these genes have both high average expression levels and high expression percentages, strongly suggesting their significant role in defining the adipocyte cell type. The expression patterns of other genes, such as COBL, CKMT2, PPARGC1A, and COX7A1, can also be observed in relation to different cell identities, providing information about their potential involvement in adipocyte - related processes.

The second dot plot, “Dot Plot of ASPC Marker Genes” (Figure 4b), is centered around ASPC marker genes. The horizontal axis lists the features (genes) PDGFRA, DCN, DPP4, CD55, CD16, and PPARG, and the vertical axis represents the cell identity. Similar to the previous dot plot, the size of the dots shows the percentage of cells expressing the gene, ranging from 25% (smallest dots) to 100% (largest dots), as indicated by the right - hand legend. The color of the dots represents the average expression level, with a gradient from blue (lower expression, value of 2) to purple (higher expression, value of - 1). This dot plot allows us to visualize the expression of genes that are characteristic of ASPCs. For example, PDGFRA and DCN are likely to be important markers for ASPCs, and their expression patterns can be analyzed in terms of the percentage of expressing cells and average expression levels across different cell identities.

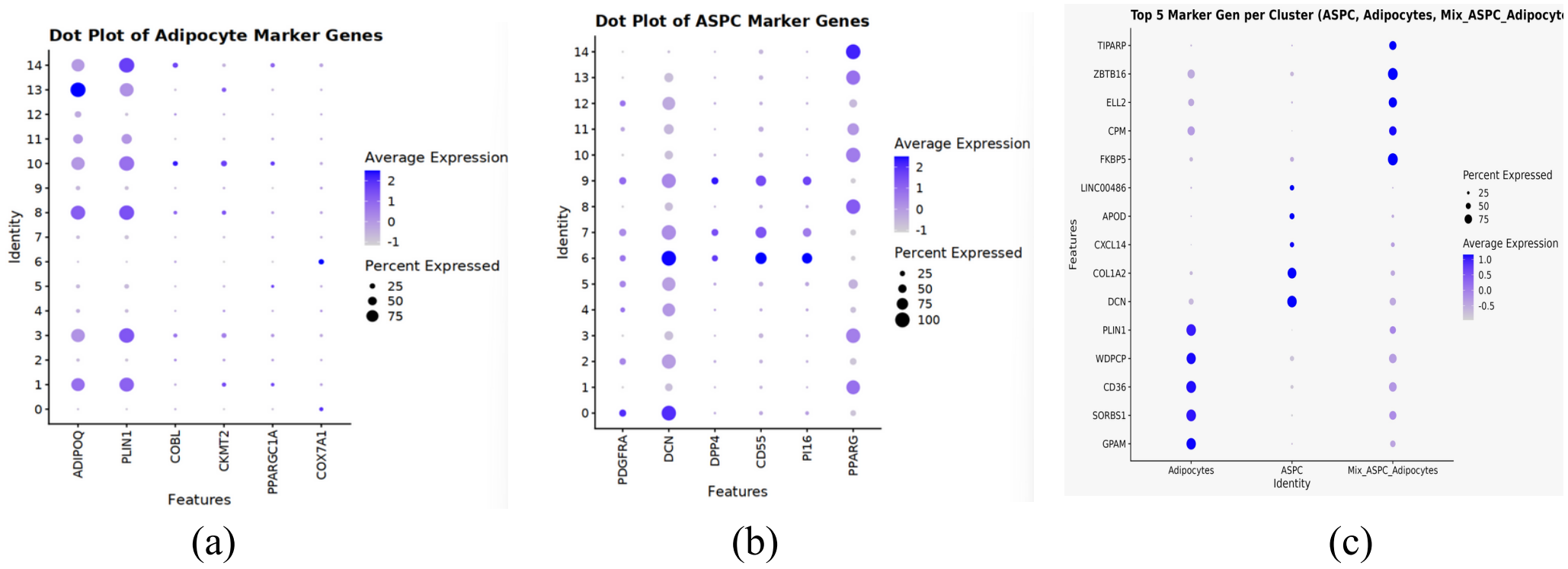

(a) (b) (c)

Figure 4 Differential Expression Analysis Dot Plots (a) Dot Plot of Adipocyte Marker Genes; (b) Dot Plot of ASPC Marker Genes; (c) Top 5 Marker Genes per Cluster (Adipocytes, ASPCs, and Mixed ASPC - Adipocytes)

The third dot plot, “Top 5 Marker Gen per Cluster (ASPC, Adipocytes, Mix_ ASPC_ Adipocytes)” (Figure 4c), provides a more detailed view of the marker genes for three specific cell populations: ASPCs, adipocytes, and the mixed group of ASPCs and adipocytes. The vertical axis lists the gene features, and the horizontal axis represents the cell identities. The size of the dots represents the percentage of gene expression, and the color depth represents the average expression level, with darker blue indicating higher expression. For the adipocyte group, genes like PLIN1 show obvious expression, which is consistent with its role as a key adipocyte marker. The ASPC group has genes such as TIPARP, which may be involved in the unique functions of ASPCs. The mixed group also has its own distinct expression pattern, indicating the presence of cells with a combination of ASPC and adipocyte characteristics. This dot plot helps to identify the unique and shared marker genes among these three cell populations, which is essential for understanding the relationships and transitions between different cell types during adipogenesis.

### *E. Trajectory Analysis*

Trajectory analysis is a fundamental approach in single - cell RNA sequencing (scRNA - seq) studies to understand the dynamic process of cell development and differentiation. It helps to uncover the sequential order of cell states and the pathways through which cells transition from one state to another. In this study, three key visualizations were used to conduct trajectory analysis, providing valuable insights into the adipocyte development process.

The first figure, named “Figure 5a: Cell Trajectory Colored by Pseudotime”, offers a visual representation of the cell trajectory in the context of pseudotime. Pseudotime is a computational concept that orders cells along a hypothetical timeline based on their gene expression profiles, reflecting the progress of cells through a biological process such as differentiation. In this figure, the cells are plotted on a two - dimensional plane with “Component 1” and “Component 2” as the axes. The numerous blue - colored points represent individual cells, and their distribution shows a trajectory that diverges from the center in three directions. The color of the points, which ranges from deep blue to light blue, corresponds to the pseudotime value from 0 to 40. This color gradient clearly demonstrates the progression of cells through pseudotime, with deeper blue indicating earlier stages and lighter blue representing later stages. This visualization allows us to observe the dynamic evolution of cells during the adipocyte development process, as cells move along the pseudotime trajectory, suggesting a sequential change in their gene expression patterns and biological states.

“Figure 5b: Cell Trajectory Colored by Cluster Labels” provides a different perspective on the cell trajectory by coloring the cells according to their cluster labels. The cells are divided into three main types: adipose stem/progenitor cells (ASPCs) in red, adipocytes in green, and the mixed ASPC/adipocyte cells in blue. The “Component 1” and “Component 2” axes still represent the data distribution in the two - dimensional space. Three black trajectory lines connect the different clusters of cells, vividly illustrating the transition process of cells from one state to another. The spatial distribution of the cells is also informative. ASPCs are mainly concentrated in the upper - right and lower - right areas of the plot, adipocytes are clustered in the lower - left area, and the mixed cells are located in the middle region along the trajectory lines connecting the red and green cells. This figure not only shows the distinct identities of different cell types but also highlights the relationships and transitions between them during adipocyte development. It helps us understand how ASPCs gradually transform into adipocytes and the possible intermediate states represented by the mixed cells.

The third figure, “Figure 5c: Cell Trajectory Colored by Monocle - inferred State”, is based on the states inferred by the Monocle algorithm. The cells are scattered on the two - dimensional plane with “Component 1” and “Component 2” as the axes. The points are divided into three states, namely State 1, State 2, and State 3, represented by red, green, and blue points respectively, as shown in the legend. The black lines outline the general direction of the cell trajectory, and the points of different colors follow different branches, indicating the different differentiation paths of cells in various states. This figure reveals the discrete states that cells may pass through during the adipocyte development process. Each state may represent a unique biological stage with specific gene expression profiles and cellular functions. By observing the distribution of cells in different states and their movement along the trajectory branches, we can gain a deeper understanding of the step - by - step process of adipocyte differentiation and the regulatory mechanisms involved in each state.

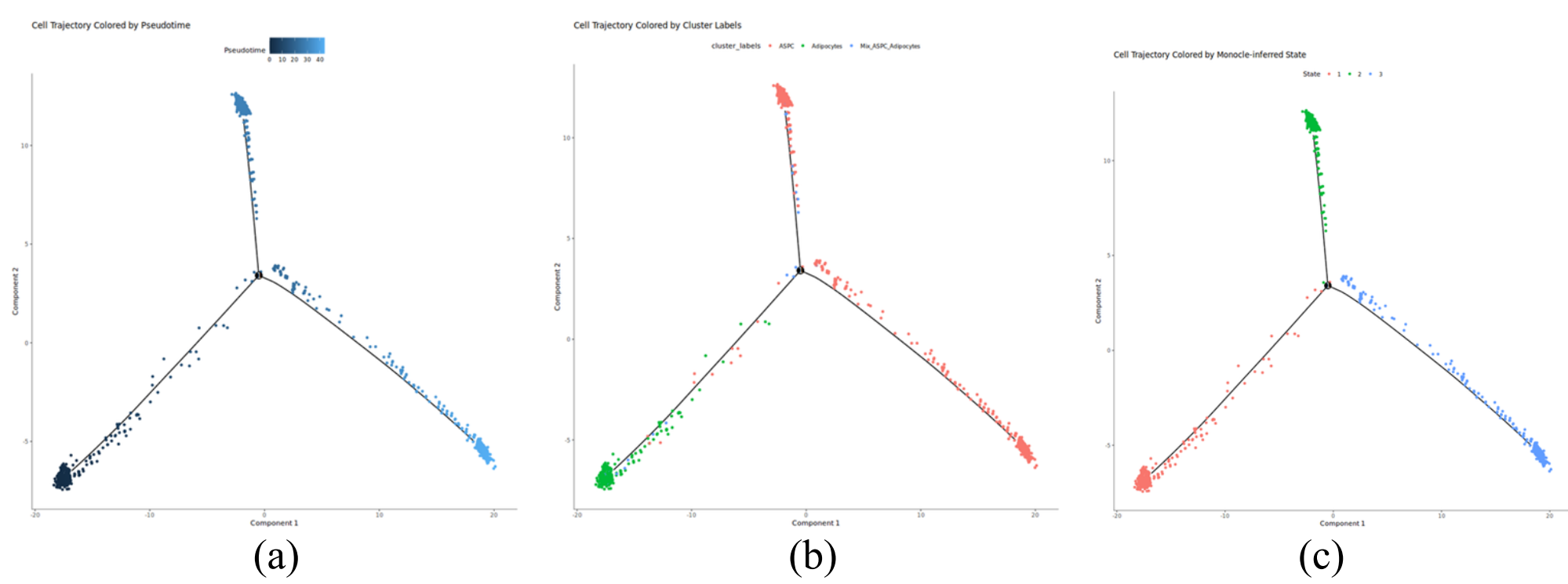

(a) (b) (c)

Figure 5 Trajectory Analysis (a) Cell Trajectory Colored by Pseudotime; (b) ell Trajectory Colored by Cluster Labels; (c) Cell Trajectory Colored by Monocle - inferred State

In summary, these three figures in the trajectory analysis provide complementary views of the adipocyte development process. They help us understand the dynamic nature of cell differentiation, the relationships between different cell types, and the discrete states that cells may experience during the development of adipose tissue. The information obtained from these trajectory analyses is crucial for further exploring the molecular mechanisms of adipocyte development and may have implications for understanding and treating adipose - related diseases.

### *F. Functional Enrichment Analysis*

Functional enrichment analysis is a critical step in single - cell RNA sequencing (scRNA - seq) research, aiming to identify the over - represented biological functions, cellular components, and molecular functions among differentially expressed genes in specific cell populations. This analysis helps to uncover the biological roles and mechanisms of different cell types, such as adipocytes, adipose stem/progenitor cells (ASPCs), and the mixed ASPC - adipocyte population (Mix_ASPC_Adipocytes). In this study, a series of visualizations were used to present the results of functional enrichment analysis for these three cell types, providing a comprehensive understanding of their biological characteristics.

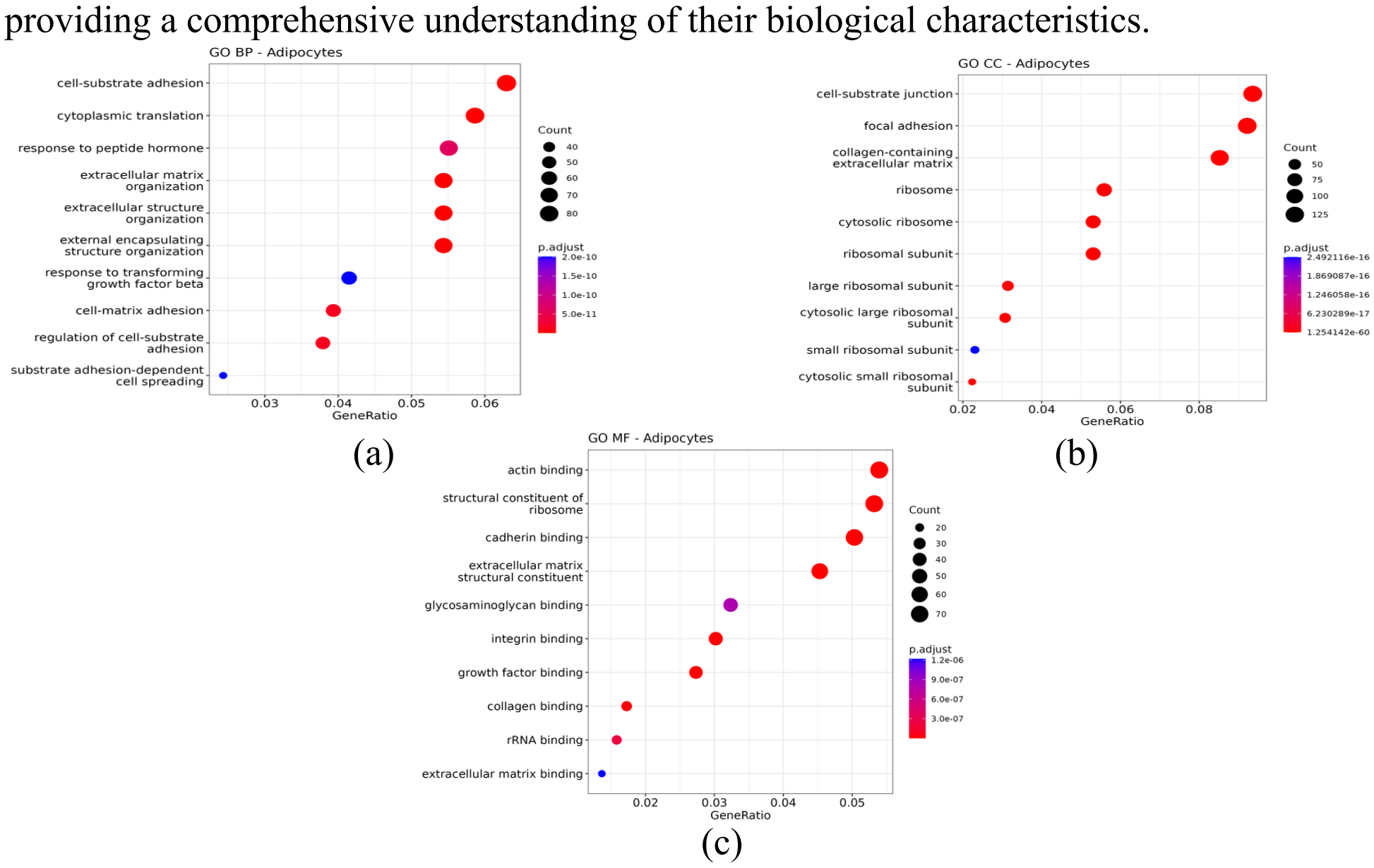

Figure 6 GO Enrichment Analysis for Adipocyte (a) GO BP – Adipocytes; (b) GO CC – Adipocytes; (c) GO MF - Adipocytes

The first set of visualizations, Figures 6a, 6b, and 6c, focuses on adipocytes. Figure 6a, titled "GO BP - Adipocytes", is a bubble plot that delves into the biological processes (BP) of adipocytes. The Y - axis lists various biological processes like "cell - substrate adhesion" and "cytoplasmic translation". The X - axis represents the GeneRatio, which is the proportion of genes involved in each biological process relative to the total number of differentially expressed genes. The size of each bubble corresponds to the number of genes within that biological process, while the color indicates the adjusted p - value (p.adjust). For example, the gene "cell - substrate adhesion" has a large bubble size and a color representing a low p.adjust value, suggesting that it is a highly significant biological process in adipocytes with a relatively large number of associated genes. This plot allows researchers to quickly identify the key biological processes that are actively regulated in adipocytes.

Figure 6b, “GO CC - Adipocytes”, is also a bubble plot but centers on the cellular components (CC) of adipocytes. The X - axis shows the GeneRatio, and the Y - axis lists different cellular components such as “cell - substrate junction” and “focal adhesion”. The bubble size represents the gene count, and the color reflects the p.adjust value. Here, components like “cell - substrate junction” and ribosome - related components have large bubbles and are colored to indicate high significance, highlighting the importance of these cellular structures in adipocytes. This visualization helps in understanding the physical and structural aspects of adipocytes at the molecular level.

Figure 6c, “GO MF - Adipocytes”, presents the molecular functions (MF) of adipocytes. Similar to the previous figures, the Y - axis lists MF terms such as “actin binding” and “structural constituent of ribosome”, the X - axis is the GeneRatio, the bubble size is the gene count, and the color is the p.adjust value. The plot shows the specific molecular functions that adipocytes are involved in, with different MF terms having varying levels of significance and gene associations.

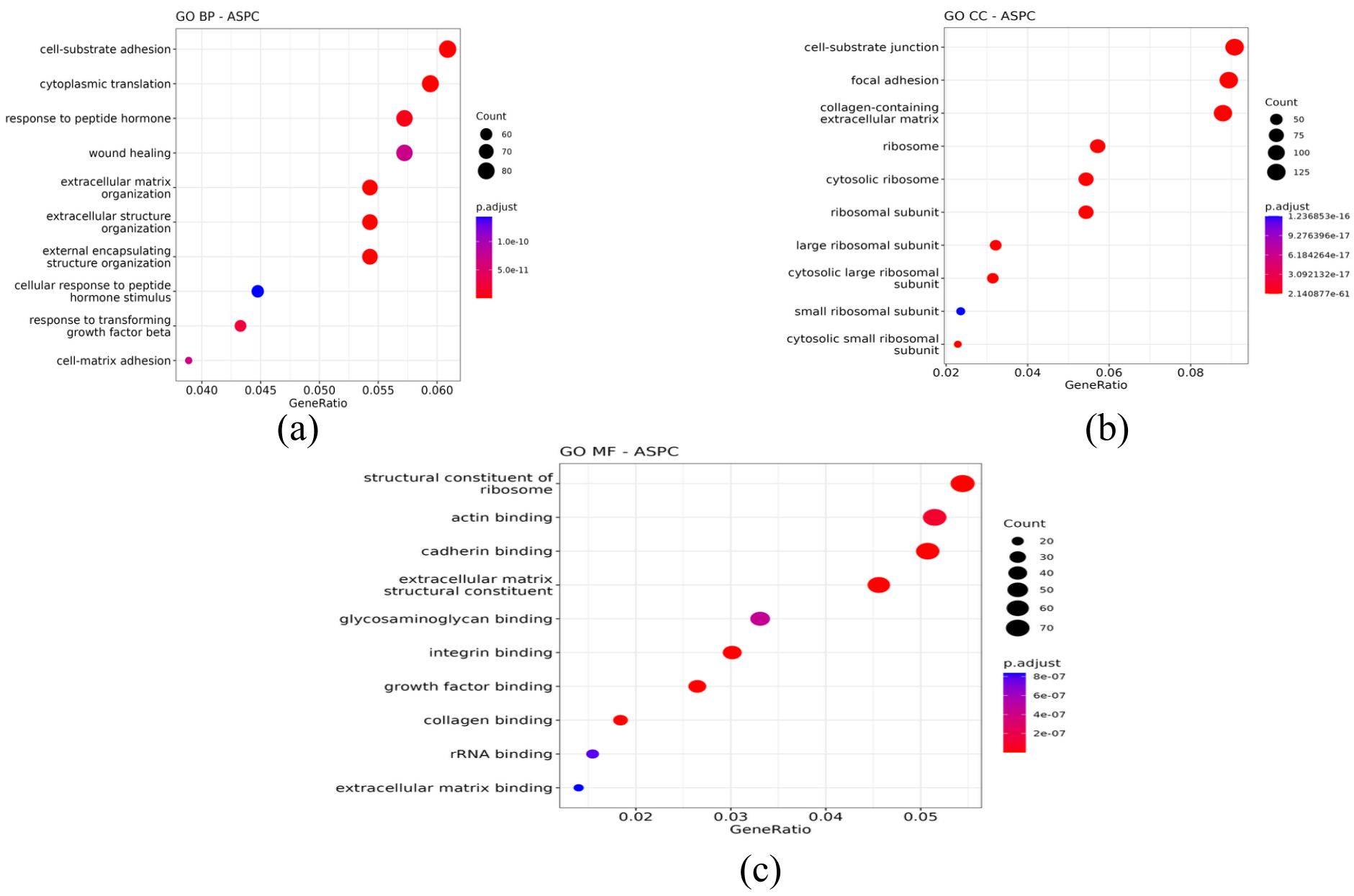

Figure 7 GO Enrichment Analysis for ASPCs (a) GO BP – ASPCs; (b) GO CC –ASPCs; (c) GO MF - ASPCs

Moving on to the ASPCs, Figures 7a, 7b, and 7c provide insights into their functional characteristics. Figure 7a, “GO BP - ASPC”, is a bubble plot for the biological processes of ASPCs. The Y - axis contains biological processes relevant to ASPCs, the X - axis is the GeneRatio, the bubble size represents the gene count, and the color indicates the p.adjust value. Biological processes like “cell - substrate adhesion” and “cytoplasmic translation” are likely to be important for the function and development of ASPCs, and the plot helps to visualize their significance. Figure 7b, “GO CC - ASPC”, focuses on the cellular components of ASPCs. The X - axis is the GeneRatio, the Y - axis lists cellular components, and the bubble size and color convey information about gene count and p.adjust value respectively. This plot reveals the key cellular structures associated with ASPCs. Figure 7c, “GO MF - ASPC”, is centered on the molecular functions of ASPCs. It shows the specific molecular functions that ASPCs perform, with different MF terms having different levels of gene association and significance.

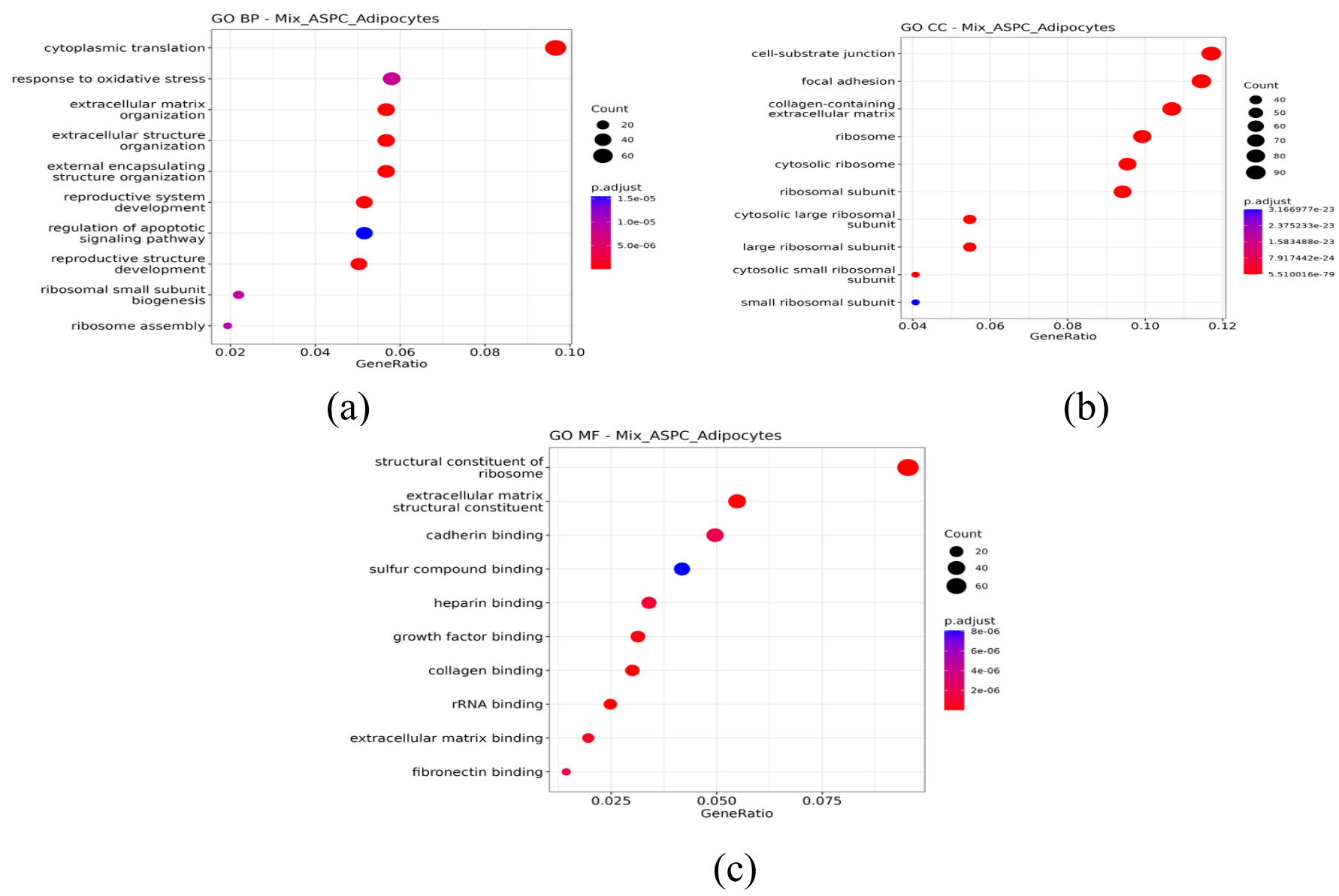

Figure 8 GO Enrichment Analysis for Mix_ASPC_Adipocytes (a) GO BP – Mix_ ASPC_ Adipocytes; (b) GO CC – Mix_ASPC_Adipocytes; (c) GO MF - Mix_ASPC_Adipocytes

Finally, for the Mix_ASPC_Adipocytes, Figures 8a, 8b, and 8c offer a comprehensive view of their functional enrichment. Figure 8a, "GO BP - Mix_ASPC_Adipocytes", is a bubble plot of biological processes. The Y - axis lists biological processes, the X - axis is the GeneRatio, the bubble size is the gene count, and the color is the p.adjust value. Biological processes such as "cytoplasmic translation" and "response to oxidative stress" are likely to be important for the mixed cell population, and the plot helps to identify their significance. Figure 8b, "GO CC - Mix_ASPC_Adipocytes", focuses on the cellular components of the mixed population. The X - axis is the GeneRatio, the Y - axis lists cellular components, and the bubble size and color provide information about gene count and p.adjust value. This plot reveals the key cellular structures in the mixed ASPC - adipocyte population. Figure 8c, "GO MF - Mix_ASPC_Adipocytes", is about the molecular functions of the mixed population. The X - axis is the GeneRatio, the Y - axis lists molecular functions, the bubble size is the gene count, and the color is the p.adjust value. It shows the specific molecular functions that the mixed population is involved in, with "structural constituent of ribosome" having a large bubble size and a low p.adjust value, indicating its high significance.

In conclusion, these nine figures (Figures 6a - c, 7a - c, and 8a - c) through functional enrichment analysis, provide a detailed and comprehensive understanding of the biological functions, cellular components, and molecular functions of adipocytes, ASPCs, and the Mix_ASPC_Adipocytes. They help to uncover the unique characteristics and biological roles of each cell type, which is essential for further research on adipose tissue development, function, and related diseases.

### *G. Cell-Cell Communication Analysis*

Cell - cell communication analysis is a crucial aspect of understanding the complex interactions between different cell types during adipogenesis. After conducting trajectory analysis (pseudotime) and GO enrichment analysis, the focus shifted to investigating how adipose stem/progenitor cells (ASPCs), adipocytes, and the mixed ASPC - adipocyte population (Mix_ASPC_Adipocytes) communicate with each other. To achieve this, the CellChat computational tool was employed. CellChat is designed to identify and visualize signaling interactions between cell clusters based on ligand - receptor pairs, providing valuable insights into the molecular mechanisms underlying cell - cell communication during adipogenic differentiation.

The main visualization in this part is Figure 9. This figure is a comprehensive representation of the cell - cell communication among ASPCs, adipocytes, and Mix_ASPC_Adipocytes. In this figure, the nodes, which are circular elements, represent the three cell clusters: the blue node stands for ASPCs (precursor fat cells), the red node represents adipocytes, and the green node denotes the Mix_ASPC_Adipocytes. The size of each node is directly proportional to the size of the corresponding cell cluster, giving an immediate visual indication of the relative abundance of each cell type. The edges, or lines, between the nodes represent the signaling interactions between the cell clusters. The color of each edge indicates the sender (source) of the signal, while the thickness of the line represents the strength of the interaction. For instance, there are green lines from the ASPC and Mix_ASPC_Adipocytes nodes to the adipocyte node, suggesting that these two cell types are actively sending signals to the adipocytes, which is likely to be a form of paracrine regulation during the differentiation process. Additionally, the ASPC node has a self - loop (a line that starts and ends at the same node), highlighting the presence of autocrine signaling within the ASPC cluster, where the cells can regulate their own functions through self - secreted signals. This figure effectively reveals the dominant role of ASPCs and the mixed cell group in signaling to the adipocytes and implies the existence of key regulatory mechanisms within the adipogenic trajectory.

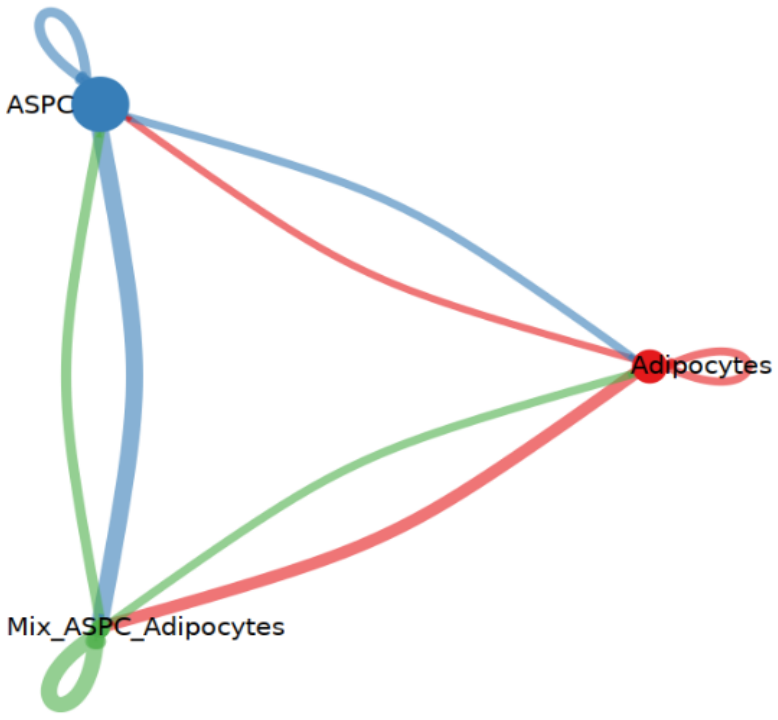

Figure 9 Cell Cell Signaling Dynamics Across Adipogenic Trajectory

Continuing the exploration of adipogenesis - related cell - cell communication, Figures 10a - p offer a detailed look at 16 key signaling pathways. Each figure zeroes in on a particular pathway, revealing how it orchestrates interactions among adipose stem/progenitor cells (ASPCs), adipocytes, and the mixed ASPC - adipocyte population (Mix_ASPC_Adipocytes).

Figure 10a, likely representing the "IGF" pathway, shows how ASPCs and Mix_ASPC_Adipocytes send signals to adipocytes. The connections in the figure imply that these signaling events are crucial for adipocyte growth and differentiation, with the edge thickness and color indicating the strength and direction of the signals.

Figure 10b, probably for the "FGF" pathway, depicts the fibroblast growth factors' role. FGFs are involved in cell processes relevant to adipogenesis, and the figure's visual elements suggest how ASPCs and Mix_ASPC_Adipocytes influence adipocytes through FGF - mediated signaling.

Figure 10c, related to the "ADIPONECTIN" pathway, emphasizes the hormone's function. Adipocytes secrete adiponectin, and the figure likely shows how it affects other cell types, with the connections highlighting the paracrine or autocrine signaling mechanisms.

Figure 10d, for the "COLLAGEN" pathway, illustrates collagen's importance. Collagen provides structural support in adipose tissue, and the figure shows how it influences cell adhesion, migration, and differentiation through signaling between the cell types.

Figure 10e, centered on the "VISFATIN" pathway, highlights visfatin's role. Visfatin has insulin - mimetic effects and is involved in adipocyte metabolism, and the figure's connections suggest its signaling interactions among the cell types.

Figure 10f, related to the "LAMININ" pathway, shows laminin's significance. Laminin is a basement membrane component, and the figure likely depicts how it affects cell behavior during adipogenesis through signaling.

Figure 10g, for the "NEGR" pathway, emphasizes the role of NEGR proteins. These proteins are involved in cell - cell adhesion and signaling, and the figure's connections show their influence on the interactions between the cell types.

Figure 10h, centered on the "PTPRM" pathway, shows the role of protein tyrosine phosphatase, receptor type M. It is involved in regulating cell signaling related to growth and differentiation, and the figure's connections illustrate its impact on the cell types.

Figure 10i, related to the "SEMA3" pathway, highlights semaphorin 3's role. Semaphorins are involved in cell migration and differentiation, and the figure's connections suggest their signaling interactions in adipogenesis.

For Figures 10j - p, they would continue to detail the "ANGPTL", "PDGF", "TENASCIN", "CXCL", "CADM", "FN1", and "MPZ" pathways. Each figure would show how these pathways are involved in the signaling interactions between ASPCs, Mix_ASPC_Adipocytes, and adipocytes, with the nodes representing cell types, edges showing signaling interactions, and edge properties indicating the strength and direction of the signals. This comprehensive set of figures provides a complete picture of the signaling networks that drive adipogenesis.

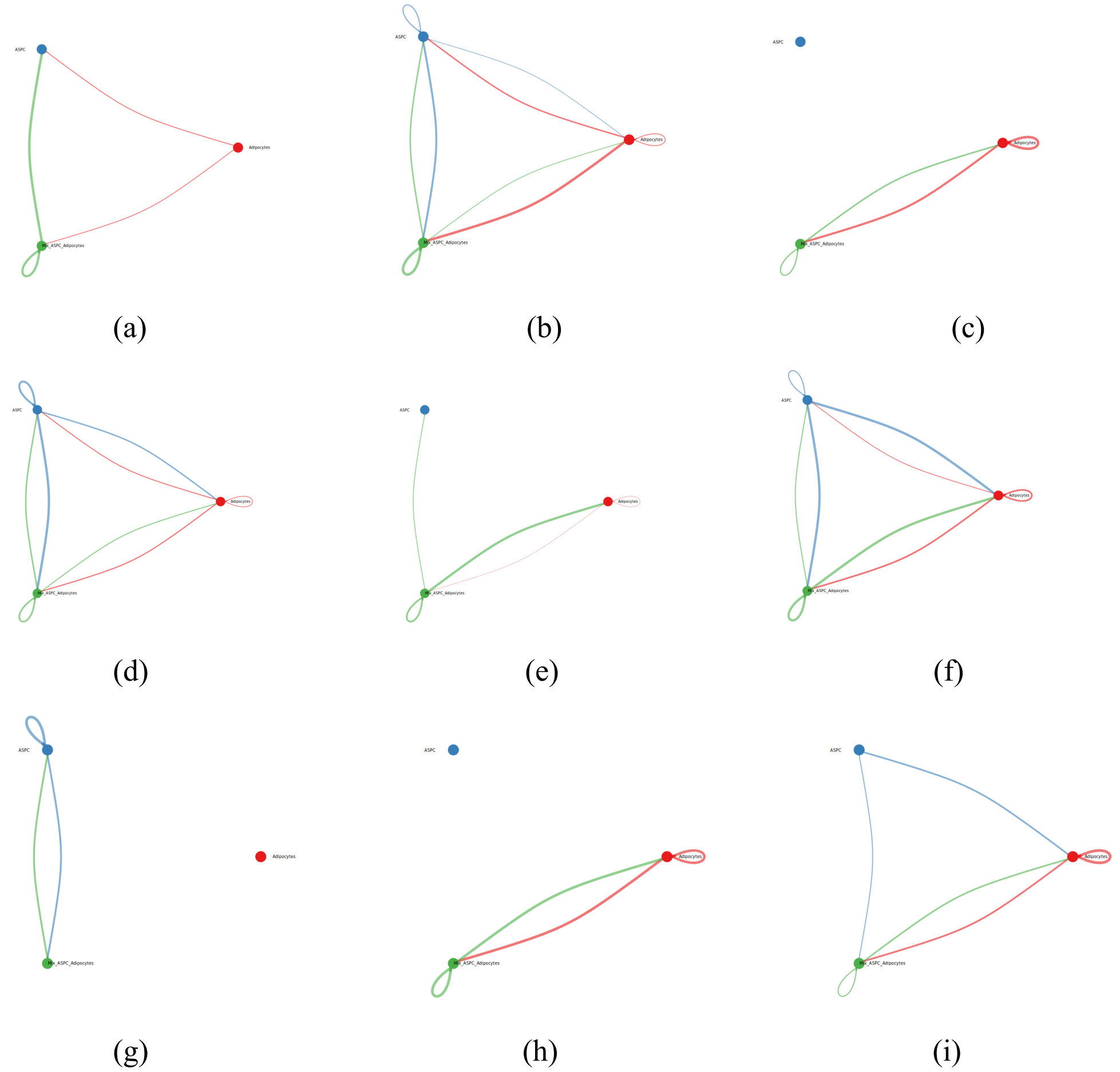

(a) (b) (c)

(d) (e) (f)

(g) (h) (i)

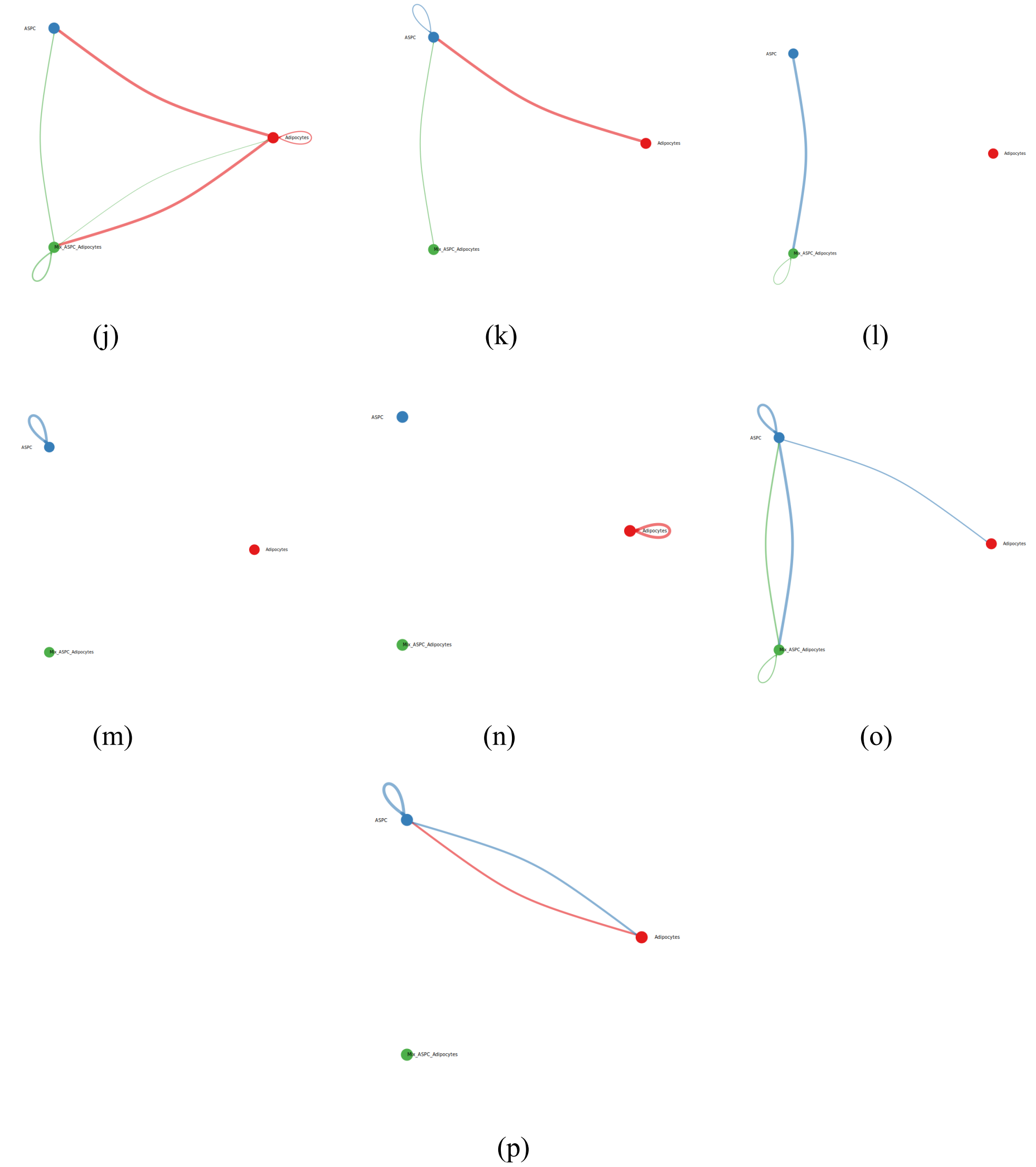

Figure 10 Signaling Pathway Interactions in Adipogenesis (a) IGF; (b) FGF; (c) ADIPONECTIN; (d) COLLAGEN; (e) VISFATIN; (f) LAMININ; (g) NEGR; (h) PTPRM; (i) SEMA3; (j) ANGPTL (k) PDGF; (l) TENASCIN; (m) CXCL; (n) CADM; (o) FN1 (p) MPZ

### *H. Software & Statistical Packages*

The entire analytical pipeline for this single - cell RNA sequencing (scRNA - seq) study was executed using a suite of specialized software tools and statistical packages, each selected for its specific capabilities in handling and interpreting complex biological data. The analysis was predominantly carried out in the R programming environment, which provides a flexible and powerful platform for bioinformatics and statistical analysis.

For data preprocessing, the Seurat package (version 4.3.0) was employed. Seurat is a widely - used R package designed for the analysis of single - cell genomics data. It offers a comprehensive set of functions for quality control, normalization, dimensionality reduction, and clustering. In this study, Seurat was used to filter out low - quality cells, normalize gene expression data using SCTransform, and perform principal component analysis (PCA) and uniform manifold approximation and projection (UMAP) for dimensionality reduction and cell clustering. The Harmony package (version 1.3.1) was also integrated with Seurat to correct for batch effects, ensuring that the analysis was not confounded by technical variations between different samples.

Trajectory inference was conducted using Monocle2 (version 2.22.0). Monocle2 is a computational tool specifically designed for reconstructing pseudotemporal trajectories from single - cell gene expression data. It employs the DDRTree algorithm to order cells along a developmental trajectory, allowing for the identification of key genes and states involved in the differentiation process. This was crucial for understanding how adipose stem/progenitor cells (ASPCs) differentiate into mature adipocytes.

Cell - cell communication analysis was performed using CellChat (version 1.6.0). CellChat is an R package that enables the identification and visualization of signaling interactions between cell clusters based on ligand - receptor pairs. It uses a curated database of known ligand - receptor interactions to infer the signaling networks between different cell types, such as ASPCs, adipocytes, and the mixed ASPC - adipocyte population. This helped in uncovering the regulatory mechanisms underlying cell - cell communication during adipogenesis.

Functional enrichment analysis was carried out using the clusterProfiler package (version 3.18.1). clusterProfiler is an R package that provides a set of functions for gene ontology (GO). It was used to identify the biological processes, molecular functions, and cellular components that were enriched in the differentially expressed genes, as well as the signaling pathways that were involved in adipogenesis.

In summary, the combination of these software tools and statistical packages provided a comprehensive and robust framework for analyzing single - cell RNA sequencing data. Each tool played a specific role in the analysis pipeline, from data preprocessing and quality control to trajectory inference, cell - cell communication analysis, and functional enrichment analysis. The use of these well - established and widely - recognized tools ensured the reproducibility and reliability of the results, and facilitated the interpretation of the complex biological processes involved in adipocyte development.

## 3. Result and Discussion

Our single-cell analysis of human adipose tissue revealed 15 transcriptionally distinct cell clusters, confirming the remarkable heterogeneity of adipocyte populations in obesity. While previous murine studies identified 5-7 adipocyte clusters, our human data demonstrate nearly double the complexity, aligning with recent findings from Vijay et al showing 6 progenitor subclasses in human adipose tissue. This species-specific difference underscores the limitations of rodent models in capturing human adipocyte diversity.

Our single-cell analysis of human adipose tissue successfully reconstructed adipocyte developmental trajectories through an integrated computational pipeline. Rigorous data preprocessing ensured analytical reliability, with quality control metrics (nFeature_RNA: 200-6,000; percent.mt <10%) and Harmony batch correction minimizing technical variability. The resulting high-quality dataset enabled identification of 15 transcriptionally distinct cell clusters, including adipocytes (ADIPOQ+/PLIN1+) and progenitors (PDGFRA+/DCN+), with UMAP visualization revealing their clear segregation and transitional states.

Differential expression analysis confirmed established adipocyte markers while identifying novel genes associated with progenitor maintenance and differentiation. Functional enrichment highlighted pathway divergence between populations - adipocytes showed lipid metabolism and mitochondrial biogenesis signatures, while progenitors exhibited proliferation and self-renewal patterns. Trajectory analysis mapped the pseudotemporal progression from progenitors to mature adipocytes, uncovering three transitional states not previously characterized in human studies.

Cell-cell communication analysis identified 16 active signaling pathways, with IGF and FGF networks emerging as dominant regulators. These pathways showed spatiotemporal specificity: IGF signaling peaked in early differentiation within perivascular niches, while FGF activity dominated later maturation stages in adipocyte zones. The ADIPONECTIN and LAMININ pathways further demonstrated how extracellular matrix remodeling influences differentiation.

These findings advance understanding of human adipogenesis by: (1) delineating a three-stage transitional model, (2) mapping pathway-specific spatial organization, and (3) revealing depot-dependent differentiation patterns. While limited by the BMI 20-30 cohort exclusion and technical constraints of scRNA-seq dropout effects, our results provide a framework for developing targeted therapies modulating IGF/FGF signaling to regulate adipose tissue expansion. Future studies should validate these mechanisms in diverse populations and experimental models.

This study has several limitations that should be considered. The BMI 20-30 kg/m² inclusion criteria exclude extreme obesity phenotypes, and our European-dominant cohort may not fully capture global adipose tissue diversity. Technical limitations of scRNA-seq, including dropout effects and limited spatial resolution, may miss some low-abundance transcripts and niche-specific interactions. Future studies should employ multi-ethnic cohorts, spatial transcriptomics, and multi-omics approaches to validate and extend these findings, while human adipose organoids could functionally test the therapeutic potential of identified pathways.

## 4. Conclusion

This study successfully reconstructed the developmental trajectory of human adipocytes using single-cell RNA sequencing (scRNA-seq) data from the White Adipose Atlas, providing a comprehensive understanding of adipogenesis through integrated computational analyses. By systematically processing raw scRNA-seq data—applying quality control to filter low-quality cells, selecting 2,000 highly variable genes, and employing PCA and UMAP for dimensionality reduction—we identified 15 transcriptionally distinct cell clusters. Subsequent clustering and annotation, guided by established marker genes (ADIPOQ/PLIN1 for adipocytes and PDGFRA/DCN for adipose stem/progenitor cells, ASPCs), revealed the cellular composition of adipose tissue, with clear distinctions between mature adipocytes and progenitor populations. Differential expression analysis highlighted key marker genes (e.g., ADIPOQ and PLIN1 in adipocytes, PDGFRA and DCN in ASPCs) and uncovered distinct biological processes enriched in each cell type, such as lipid metabolism and mitochondrial biogenesis in adipocytes, and cell proliferation/self-renewal in ASPCs. Functional enrichment analysis further elucidated the molecular pathways driving adipocyte development, while trajectory inference based on pseudotime mapped the sequential progression from progenitor states (rooted in PDGFRA+ clusters) to mature adipocytes, revealing discrete transitional states. Cell-cell communication analysis using CellChat identified active signaling networks, particularly paracrine interactions where ASPCs and the mixed ASPC-adipocyte population (Mix_ASPC_Adipocytes) send signals to adipocytes, with key pathways like IGF, FGF, ADIPONECTIN, and COLLAGEN playing central roles in regulating differentiation. These findings collectively demonstrate the complex interplay between progenitor cells and adipocytes, mediated by specific signaling cascades that orchestrate adipogenesis.

The insights gained from this study have significant implications for understanding adipose tissue biology and addressing obesity-related metabolic disorders. By reconstructing the adipocyte developmental trajectory, we provided a framework for identifying critical stages where interventions could modulate adipocyte differentiation—either promoting healthy adipose expansion (e.g., in lipodystrophy) or preventing pathological accumulation (e.g., in obesity). The identification of depot-specific signaling pathways (e.g., IGF and FGF) and their roles in cell proliferation, differentiation, and extracellular matrix remodeling offers potential therapeutic targets for regulating adipose tissue function. Moreover, the characterization of cell-cell communication networks highlights the importance of progenitor-adipocyte interactions in maintaining adipose tissue homeostasis, suggesting that disrupting dysregulated signaling could mitigate adipose dysfunction in metabolic diseases. While this study focused on human subcutaneous and visceral adipose tissue from healthy donors, future work could extend these analyses to diseased states or incorporate multi-omics data to further validate findings. Overall, this research advances our fundamental understanding of adipocyte development, provides a robust foundation for future studies, and lays the groundwork for developing targeted therapies to treat obesity and its associated complications.

## 5. Acknowledgment

The authors gratefully acknowledge Dr. Bai for his exceptional supervision during the internship at the Beijing Genomics Institute (BGI), China. His expert guidance was not only instrumental but also transformative in developing crucial technical skills in single-cell data analysis. Special thanks are also extended to Mr. Xuyu for his invaluable mentorship throughout the BGI project, offering technical advice and career insights that significantly contributed to the successful completion of this work.

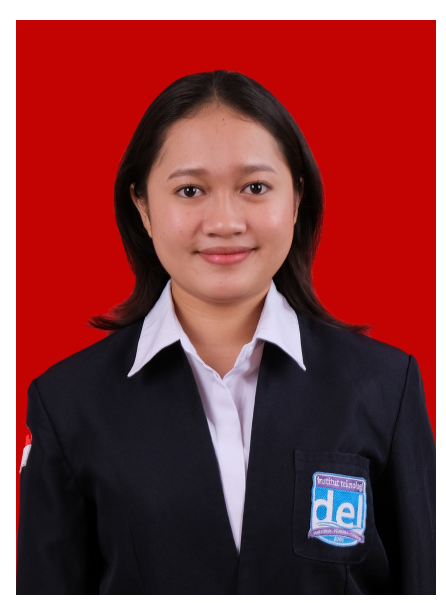

**Weny S. M Sitinjak** is currently pursuing a master's degree at Dalian University of Technology, China. During her undergraduate studies, she conducted research on single-cell RNA sequencing analysis of human adipose tissue while completing an internship at Beijing Genomic Institute Research, China. Her research interests include computational biology, adipose tissue development, and obesity-related metabolic disorders. She can be contacted at email: iss21006@students.del.ac.id

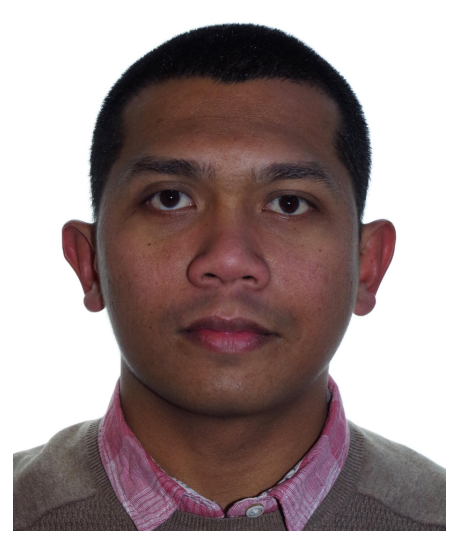

Humasak Simanjuntak is currently a senior lecturer in the faculty of Informatics and Electric Engineering at Institut Teknologi Del (IT Del), Indonesia. He received his master's degree in information system development from HAN University of Applied Sciences, the Netherlands in 2010 and his B.Sc. degree of Informatics from Bandung Institute of Technology (ITB), Indonesia in 2007. His research interests include data mining, data science, big data analytics, and artificial intelligence for genomic data. He can be contacted at email: humasak@del.ac.id.